\documentclass[aps,pra,reprint,superscriptaddress,showpacs]{revtex4-1}
\usepackage{amssymb,graphicx,epsfig,threeparttable}
\usepackage{color}
\usepackage[colorlinks=true,linkcolor=blue]{hyperref}

\begin{document}

\title{Using non-Markovian measures to evaluate quantum master equations for photosynthesis}
\author{Hong-Bin Chen}
\affiliation{Department of Physics and National Center for Theoretical Sciences, National Cheng Kung University, Tainan 701, Taiwan}
\author{Neill Lambert}
\affiliation{CEMS, RIKEN, Wako-shi, Saitama 351-0198, Japan}
\author{Yuan-Chung Cheng}
\email{yuanchung@ntu.edu.tw}
\affiliation{Department of Chemistry and Center for Quantum Science and Engineering, National Taiwan University, Taipei 106, Taiwan}
\author{Yueh-Nan Chen}
\email{yuehnan@mail.ncku.edu.tw}
\affiliation{Department of Physics and National Center for Theoretical Sciences, National Cheng Kung University, Tainan 701, Taiwan}
\author{Franco Nori}
\affiliation{CEMS, RIKEN, Wako-shi, Saitama 351-0198, Japan}
\affiliation{Physics Department, University of Michigan, Ann Arbor, Michigan 48109-1040, USA}
\date{\today}

\begin{abstract}
When dealing with system-reservoir interactions in an open quantum system, such as a photosynthetic light-harvesting complex, approximations
are usually made to obtain the dynamics of the system. One question immediately arises: how good are these approximations, and in what ways
can we evaluate them? Here, we propose to use entanglement and a measure of non-Markovianity as benchmarks for the deviation of approximate
methods from exact results. We apply two frequently-used perturbative but non-Markovian approximations to a photosynthetic dimer model and
compare their results with that of the numerically-exact hierarchy equation of motion (HEOM). This enables us to explore both entanglement
and non-Markovianity measures as means to reveal how the approximations either overestimate or underestimate memory effects and quantum
coherence.  In addition, we show that both the approximate and exact results suggest that non-Markonivity can, counter-intuitively, increase
with temperature, and with the coupling to the environment.
\end{abstract}

\maketitle

\section{INTRODUCTION}

Modelling and understanding the non-equilibrium dynamics of open quantum systems is a ubiquitous problem in physics, chemistry and biology
\cite{yuanchun_review_article,yuehnan_nat_phy_2013}. 
In such systems, the environment is usually composed
of huge numbers of microscopic constituents, an exact description of which is challenging. One can invoke intensive computational techniques,
such as path-integral formalisms \cite{leggett_rev,weiss_textbook,grabert_phys_rep,grifoni_phys_rep}, Monte Carlo algorithms \cite{egger_pre},
the hierarchy equations of motion (HEOM) \cite{tanimura_pra,tanimura_jpsj_1989,tanimura_jpsj_2006}, the reaction-coordinate method \cite{nazir_pra_2014,garg_jcp_1985}
and others, to explicitly and exactly propagate the quantum state of a complete system-environment model. However, a common drawback
of these exact numerical solutions is their demanding computational resource requirements, which can scale badly depending on the
spectral density of the environment being modelled, the number of independent baths the system is coupled to, or the complexity of
the system itself.

To simplify the problem and gain useful physical insight, approximations are usually made to reduce the system dynamics to that of a
relatively few degrees of freedom. In that regard, much effort has been devoted to develop quantum master equations (QMEs) which describe
these reduced degrees of freedom in various limits. Redfield theory \cite{redfield_1965} provides one with QME based on (together with
a secular approximation) a second-order perturbation approximation in the system-environment coupling. For the strong-coupling limit,
F\"{o}rster theory \cite{foerster_nws_1946,foerster_annphys_1948,foerster_rrs_1960}, adopts a diffusion-rate equation \cite{haenggi_rew_mod_phys_1990}
to describe the incoherent transport phenomenon. Nevertheless, these conventional Markovian QME treatments cannot capture the memory effects of the bath.

In order to take into account the memory effects, many attempts at improving these Markovian QMEs have been made. The second-order
time-convolution (TC2) \cite{arimitsu_tc} equation is known as a chronological-ordering prescription \cite{mukamel_cop} or time-nonlocal
equation \cite{Meier:1999tj,kleinekathoefer_jcp}. It is a direct generalization of Redfield QME without performing the Markov and secular
approximations. The second-order time-local (TL2) equation is another frequently used QME, sometimes called a partial-time-ordering
prescription \cite{mukamel_cop} or time-convolutionless equation. Some works suggest that TL2 shows better performance than TC2 at
numerically approximating exact results \cite{silbey_comparison}. Nevertheless, their respective domains of applicability
have not been thoroughly investigated yet.

In each QME model (TC2, TL2), certain approximations and simplifications are introduced to obtain solvable equations. To investigate
the deviation of each approximate QME model from the exact results, we first compare the explicit dynamics of these two approximative QMEs
with that of the HEOM (which is considered to be numerically exact).  We focus on the intermediate system-environment
coupling regime, which has proven to be the most challenging and relevant to the dynamics in realistic systems, and the one at which
the regime of validity of most approximations breaks down. Both approximate methods are perturbative in the system-bath coupling, but
can in principle harbor memory effects of the environment. To investigate how well they capture these effects we then utilize the concept
of the Choi-Jamio{\l}kowski isomorphism \cite{jamiolkowski,choi} to encode complete information on the dynamics of the system into the
entanglement with an ancilla. By comparing the time evolution of the entanglement between system and ancilla, and an associated measure
of non-Markovianity \cite{RHB_measure}, one can find out to what extent the memory effects and coherence predicted by each approximative
QME deviates from being numerically exact. Our results suggest that entanglement and non-Markovianity provide a useful benchmark for the
performance of such approximative treatments.

In performing this analysis we also discuss several interesting physical trends, including a counter-intuitive increase of non-Markovianity
with both temperature and with the coupling strength to the environment. We attribute this increase to an enhancement of system-environment
correlations when both the coupling and temperature are increased. Additionally, evidence from other studies
\cite{nazir_prl_2009,fleming_science_2007,scholes_nature_2010} suggests that non-Markovian environments are capable of sustaining quantum
coherence. The interplay of these factors finally results in the increase of non-Markovianity with both temperature and coupling strength
that we see in our results.

\section{THE SPIN-BOSON MODEL}

\subsection{The model}
The spin-boson model \cite{leggett_rev} is one of the most extensively studied models of open quantum systems, and is the one we employ
here. It describes a spinor-like two-state system interacting with a bosonic environment. First, let us consider this standard model,
which can be divided into three components
\begin{equation}
\widehat{H}_{\textrm{tot}}=\widehat{H}_{\textrm{sys}}+\widehat{H}_{\textrm{env}}+\widehat{H}_{\textrm{int}}.
\end{equation}
The system Hamiltonian, $\widehat{H}_{\textrm{sys}}$, is written as
\begin{equation}
\widehat{H}_{\textrm{sys}}=\frac{\hbar \omega _{0}}{2}\hat{\sigma}_{z}+J\hat{\sigma}_{x},
\end{equation}
where $J\hat{\sigma}_{x}$ is the coherent-coupling term, which enables the tunneling between the two system quantum states, labeled
as $|1\rangle$ and $|-1\rangle$, with the energy level spacing $\hbar \omega _{0}$. Usually, one adopts the delocalized basis
$|\chi _{+}\rangle $ and $|\chi _{-}\rangle$ (exciton), which is defined by the following eigenvalue problem
\begin{equation}
\left( \frac{\hbar \omega _{0}}{2}\hat{\sigma}_{z}+J\hat{\sigma}_{x}\right)
\mid \chi _{\pm }\rangle =\pm \frac{\hbar }{2}\Omega \mid \chi _{\pm
}\rangle,
\end{equation}
with $\Omega =\sqrt{\omega _{0}^{2}+4J^{2} / \hbar ^{2}}$.

The environment, $\widehat{H}_{\textrm{env}}$, is usually modelled as a large collection of harmonic oscillators
\begin{equation}
\widehat{H}_{\textrm{env}}=\sum_{\mathbf{k}}\hbar \omega _{\mathbf{k}}\hat{a}_{
\mathbf{k}}^{\dagger }\hat{a}_{\mathbf{k}},
\end{equation}
where $\hat{a}_{\mathbf{k}}^{\dagger }$ ($\hat{a}_{\mathbf{k}}$) is the creation (annihilation) operator of the environment mode
$\mathbf{k}$ with angular frequency $\omega _{\mathbf{k}}$. For simplicity, a linear system-environment coupling,
$\widehat{H}_{\textrm{int}}$, is adopted throughout this work:
\begin{equation}
\widehat{H}_{\textrm{int}}=\sum_{\mathbf{k}}\hat{\sigma}_{z}\otimes \left( \hbar g_{
\mathbf{k}}\hat{a}_{\mathbf{k}}^{\dagger }+\hbar g_{\mathbf{k}}^{\ast }
\hat{a}_{\mathbf{k}}\right) ,
\end{equation}
where $g_{\mathbf{k}}$ is the coupling constant between the environment mode $\mathbf{k}$ and the system. In most physical problems, the
details of the microscopic description of $g_{\mathbf{k}}$ are not clear, and one usually employs a spectral density function,
$J(\omega )=\sum_{\mathbf{k}}|g_{\mathbf{k}}|^{2}\delta (\omega -\omega _{\mathbf{k}})$, to characterize the coupling strength via the
reorganization energy $\lambda =\int_{0}^{\infty }J(\omega)/\omega d\omega $. The physical meaning of the spectral density function can
be understood as the density of states of the environment, weighted by the coupling strengths. Moreover the way in which the environment
modulates the dynamics of the system is described by the correlation function
\begin{equation}
G(t)=\int_{0}^{\infty }J(\omega )\left[ \coth \left(\frac{\hbar \omega }{2k_{B}T}
\right)\cos \omega t-i\sin \omega t\right] d\omega .  \label{correlation_function}
\end{equation}
The real part is related to the dissipation process, while the imaginary part corresponds to the response function.

The statistical properties of the entire system can be described by the total density matrix $\rho _{\textrm{tot}}$, which contains all
the degrees of freedom of the system and environment. If the correlation between the system and environment is negligible, the Born
approximation can be used and the total density matrix can be factorized into
\begin{equation}
\rho _{\textrm{tot}}(t)=\rho _{\textrm{sys}}(t)\otimes \rho _{\textrm{env}},
\end{equation}
where $\rho_{\textrm{sys}}(t)$ describes the dynamics of the system and $\rho_{\textrm{env}}=\exp\left[-\widehat{H}_{\textrm{env}}/k_{B}T\right]/Z$
is the environment density matrix in thermal equilibrium at temperature $T$. Here, $k_{B}$ is the Boltzmann constant and
$Z=\textrm{Tr}\exp\left[-\widehat{H}_{\textrm{env}}/k_{B}T\right]$ is the partition function.

One notes that when $\omega _{0}$, $J$, and $\lambda$ are comparable, this makes the conventional perturbative treatment unreliable.
In the following, we will adopt the two frequently-used perturbative but non-Markovian QME formalisms discussed in the introduction and
compare their results with the exact one in the intermediate-coupling regime, as they both begin to break down, and investigate ways in which to
evaluate their accuracy.

\subsection{Second-order time-convolution equation (TC2)}

For the Hamiltonian defined above, the time evolution of the system density matrix $\rho _{\textrm{tot}}(t)$ under the TC2 approximation
is expressed as
\begin{eqnarray}
&&\frac{\partial}{\partial t}\rho_{\textrm{sys}}(t)=-\frac{1}{\hbar^{2}}\textrm{Tr}_{\textrm{env}} \nonumber \\
&&\times\int_{0}^{t}\left[\widetilde{H}_{\textrm{int}}(t),\left[\widetilde{H}_{\textrm{int}}(\tau),\rho_{\textrm{sys}}(\tau)
\otimes\rho_{\textrm{env}}\right]\right]d\tau.
\label{master_equation_TC2_J_coupling}
\end{eqnarray}
The tilde symbol above an operator denotes the interaction picture with respect to $\widehat{H}_{\textrm{sys}}+\widehat{H}_{\textrm{env}}$.
The interaction Hamiltonian in terms of the delocalized basis can be expressed as
\begin{eqnarray}
\widetilde{H}_{\textrm{int}}(t)=\sum_{\mu,\nu,\mathbf{k}}&&A_{\mu,\nu}e^{i\omega_{\mu,\nu}t}|\mu\rangle\langle\nu|  \nonumber \\
&\otimes&\left(\hbar g_{\mathbf{k}}\hat{a}_{\mathbf{k}}^{\dagger}e^{i\omega_{\mathbf{k}}t}+\hbar g_{\mathbf{k}}^{\ast}\hat{a}_{\mathbf{k}}
e^{-i\omega_{\mathbf{k}}t}\right),  \label{interaction_hamiltonian}
\end{eqnarray}
where $A_{\mu,\nu}=\langle\mu|\hat{\sigma}_{z}|\nu\rangle$, and $\mu,\nu=\chi_{+},\chi_{-}$. Substituting
Eq.~(\ref{interaction_hamiltonian}) into (\ref{master_equation_TC2_J_coupling}) with the explicit expansion leads to a set of
simultaneous integrodifferential equations of the density matrix elements $\rho_{\mu,\nu}(t)$
\begin{eqnarray}
\frac{\partial}{\partial t}\left[e^{-i\omega_{\alpha,\beta}t}\rho_{\alpha,\beta}(t)\right]=
-i\omega_{\alpha,\beta}e^{-i\omega_{\alpha,\beta}t}\rho_{\alpha,\beta}(t)   \nonumber\\
+\sum_{\mu,\nu}\int_{0}^{t}f_{\mu,\nu}(t-\tau)e^{-i\omega_{\mu,\nu}\tau}\rho_{\mu,\nu}(\tau)d\tau.
\label{int_diff_TC2}
\end{eqnarray}
One notes that the memory effects are taken into account in terms of the convolution of the memory kernel $f_{\mu ,\nu }(t-\tau )$.
A detailed expression for this kernel is given in the Appendix.

To solve the simultaneous integrodifferential components of Eq.~(\ref{int_diff_TC2}), we invoke the Laplace transformation
$\mathcal{L}\{f\}:=\int_{0}^{\infty}f(t)e^{-st}dt$, and transform them into a set of algebraic equations. After carefully analyzing the
properties of the poles, the conventional residual theorem enables one to accomplish the inverse Laplace transformation and move back
from Laplace space into the time-domain.

\subsection{Second-order time-local equation (TL2)}

In the TL2 formalism, the system is considered to be sluggish, hence the bath feedback on the system dynamics can be neglected by
approximating $\rho_{\textrm{sys}}(t-\tau)\approx\exp\left[i\widehat{H}_{\textrm{sys}}\tau/\hbar\right]\rho_{\textrm{sys}}(t)\exp
\left[-i\widehat{H}_{\textrm{sys}}\tau/\hbar\right]$. This assumption is reasonable because it is impossible for a system to change
its configuration instantaneously. Consequently the system density matrix should be pulled out from the integral to obtain the
following QME
\begin{eqnarray}
&&\frac{\partial}{\partial t}\rho_{\textrm{sys}}(t)=-\frac{1}{\hbar ^{2}}\textrm{Tr}_{\textrm{env}} \nonumber \\
&&\times\int_{0}^{t}\left[\widetilde{H}_{\textrm{int}}(t),\left[\widetilde{H}_{\textrm{int}}(\tau),\rho_{\textrm{sys}}(t)
\otimes\rho_{\textrm{env}}\right]\right]d\tau .
\label{master_equation_TL2_J_coupling}
\end{eqnarray}
Similarly, substituting Eq.~(\ref{interaction_hamiltonian}) into (\ref{master_equation_TL2_J_coupling}) with the explicit expansion
leads to a set of simultaneous differential equations of the density matrix elements $\rho_{\mu,\nu}(t)$
\begin{eqnarray}
\frac{\partial}{\partial t}\left[e^{-i\omega_{\alpha,\beta}t}\rho_{\alpha,\beta}(t)\right]=
-i\omega_{\alpha,\beta}e^{-i\omega_{\alpha,\beta}t}\rho_{\alpha,\beta}(t)   \nonumber\\
+\sum_{\mu,\nu}\left(\int_{0}^{t}h_{\mu,\nu}(t-\tau)d\tau\right)e^{-i\omega_{\mu,\nu}t}\rho_{\mu,\nu}(t).
\label{int_diff_TL2}
\end{eqnarray}
The detailed expression of the memory kernel $h_{\mu ,\nu }(t-\tau )$ is given in the Appendix. It should be emphasized that although
$\rho_{\mu,\nu}(t)$ is pulled out from the integral, Eq.~(\ref{int_diff_TL2}) is capable of predicting a non-Markovian dynamics
because the time integral of $h_{\mu ,\nu }(t)$ results in time-varying coefficients in front of $\rho_{\mu,\nu}(t)$. Whether or
not such differential equations behave non-Markovianly  crucially depends on these time-varying coefficients.

\begin{figure*}[]
\includegraphics[width=\textwidth]{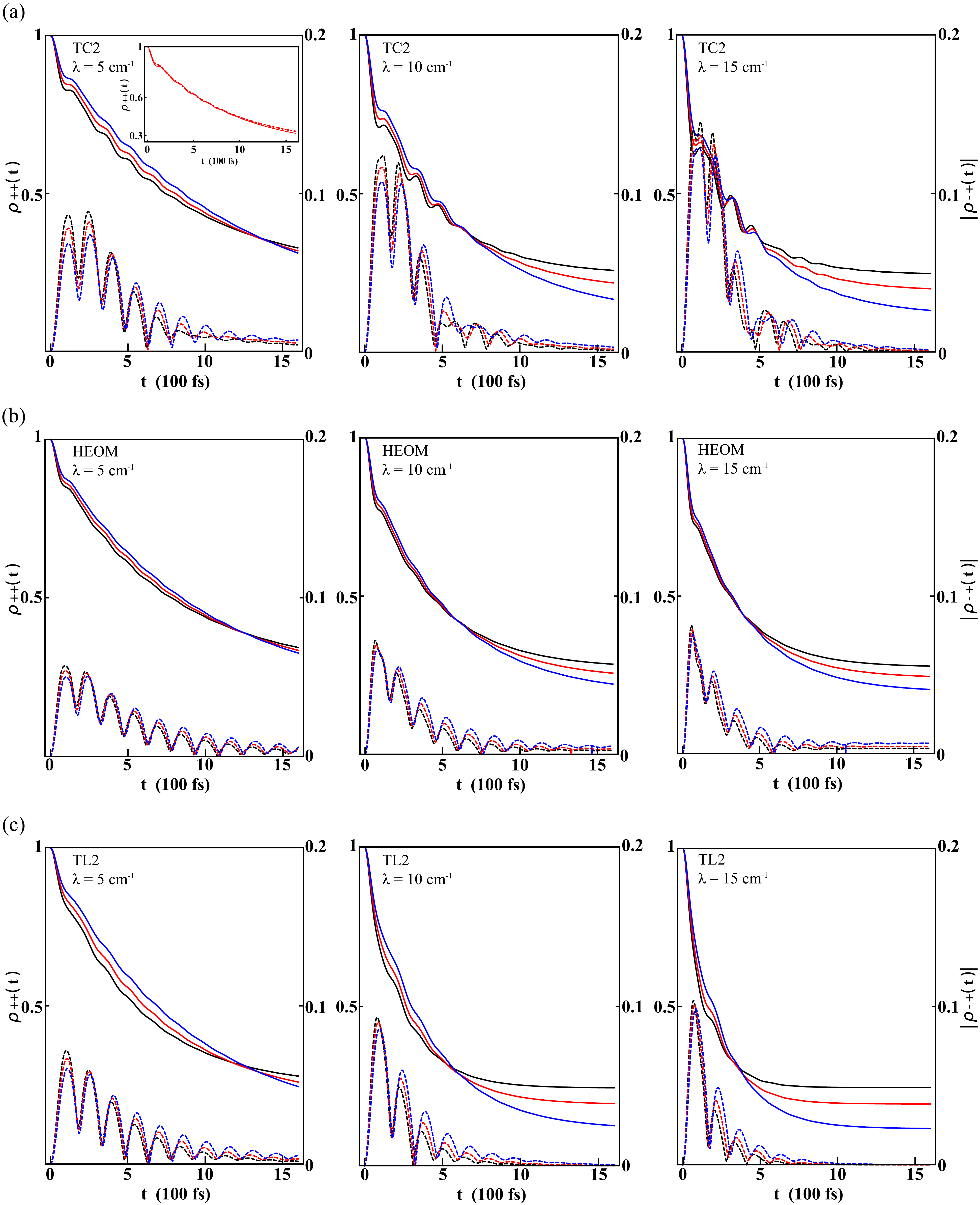}
\caption{(Color online) Time evolution of the populations $\rho_{++}(t)$ (solid) and coherence $|\rho_{-+}(t)|$ (dashed) predicted by
(a) TC2, (b) HEOM, and (c) TL2
for the spin-boson model with different values of $\lambda$ at temperatures $T=300$ K (black), $250$ K (red), and $200$ K (blue). The
other parameters are $\omega_{0}=70$ cm$^{-1}$, $J=100$ cm$^{-1}$, and $\gamma=50$ cm$^{-1}$ ($\gamma^{-1}=106$ fs). For small $\lambda$,
both QMEs yield excellent results, as expected. The inset in (a) shows the results given by TC2 (solid curve) and HEOM (dot-dashed curve)
for $\lambda=5$ cm$^{-1}$, and $T=250$ K, illustrating how they almost overlap. However, due to over-estimation of the coherence, the result
calculated from the TC2 method shows a slightly higher beating behavior in the population dynamics. In contrast, for large $\lambda$ the
population dynamics predicted by the TC2 method is in better agreement with those of the HEOM, whereas the populations given by the TL2 method
are somewhat sluggish and tend to approach thermal equilibrium a bit faster.}
\label{Dynamics}
\end{figure*}

\section{COMPARISONS WITH EXACT RESULTS}

To illustrate the differences of the approximations explicitly, we apply these two QMEs to a photosynthetic dimer model, which has
attracted considerable interest recently
\cite{yuanchun_review_article,yuehnan_nat_phy_2013,fleming_expe_pnas_2006,grondelle_pccp_2010,pachon_pccp_2012,fleming_hierar_jcp,fleming_theo_pnas_2009,hongbin_pre_2014}.
We employ the Drude-Lorentz spectral density function (the over-damped Brownian oscillator model) \cite{tanimura_jpsj_2006, mukamel_textbook},
$J(\omega)=\left(2\lambda\gamma/\pi\right)\left[\omega/\left(\omega^{2}+\gamma^{2}\right)\right]$, which has been shown to fit the
experimental data well \cite{fleming_expe_pnas_2006,grondelle_pccp_2010} and has been successfully used for a range of theoretical studies
of this type of system \cite{pachon_pccp_2012,fleming_hierar_jcp,fleming_theo_pnas_2009,hongbin_pre_2014}. As mentioned in the previous
section, the reorganization energy, $\lambda$, characterizes the coupling strength to the environment, while the quantity $\gamma$ determines
the width of the spectral density. These two parameters have considerable influence on the dynamics of the system.

In Fig.~\ref{Dynamics}, we show the system dynamics given by (a) TC2, (b) HEOM, and (c) TL2 with varying $\lambda$ and temperature $T$.
The other parameters are fixed at $\omega_{0}=70$ cm$^{-1}$, $J=100$ cm$^{-1}$, and $\gamma=50$ cm$^{-1}$ ($\gamma^{-1}=106$ fs). These
 parameters are typical in photosynthetic systems. The solid curves in each panel denote the populations
of the $|\chi_{+}\rangle$ state with temperatures $T=300$ K (black), $250$ K (red), and $200$ K (blue), respectively. It can be seen that,
at higher temperatures, the population of the $|\chi_{+}\rangle$ state transfers to the $|\chi_{-}\rangle$ state faster than at lower temperatures,
but there is always a crossing so that the thermal equilibrium population of the $|\chi_{+}\rangle$ state is larger at higher temperatures.

For small values of $\lambda$, the results of the two QME models show excellent agreement with that of the HEOM, indicating that both TC2
and TL2 perform well in the weak system-environment coupling regime and that the bath memory effect is insignificant at small $\lambda$.
Moreover, the result of TC2  completely coincides with that of the HEOM for very small couplings.
We show the comparison between TC2 (solid curve) and HEOM (dot-dashed curve) methods in the inset of Fig.~\ref{Dynamics}(a) for $\lambda=5$
cm$^{-1}$, and $T=250$ K. When $\lambda$ is increased, the TC2 population results exhibit vigorous beating and produce oscillatory curves up
to $800$ fs, which is absent in the HEOM result. We attribute these oscillations to the over-estimation of the coherence by TC2. Apart from these
beatings, the overall magnitude of the population of HEOM is quantitatively better approximated by TC2 than TL2. The TL2 model yields monotonically-decaying
population dynamics that tend to reach thermal equilibrium too rapidly. This leads to a significant over-estimation of the population relaxation rate
by TL2, especially at large $\lambda$. This over-estimation of the population relaxation rate in Redfield theory has been reported previously
\cite{fleming_redfield_jcp}, and here we gain further insight into its origin by comparing to the TC2 results.

The dashed curves in Fig.~\ref{Dynamics} denote the absolute value of the off-diagonal elements of the system density matrix, i.e., the
coherence between the $|\chi_{+}\rangle$ and $|\chi _{-}\rangle $ states. The results from the TC2 method manifestly show the over-estimation
of the coherence even if $\lambda$ is small. When $\lambda$ is increased, the over-estimation of the coherence becomes quite pronounced.
On the other hand, the coherence in the TL2 model decays more rapidly, leading to the sluggish dynamics discussed above.  In summary, the
coherence dynamics is better approximated by TL2, and the TC2 model may fail in approximating the true coherence for large $\lambda$.  However,
the overall population decay rate predicted by the TC2 is generally more correct than that of TL2. It is interesting to note that the TL2
model yields an exact QME for a pure dephasing spin-boson model (i.e. $J=0$) \cite{silbey_comparison} while the TC2 model underestimates
the pure dephasing rate, which is in line with our findings here.

\section{BENCHMARK OF APPROXIMATIVE QMEs}

\subsection{Entanglement and non-Markovianity}

In the previous section, we analyzed how the coherence terms of the two approximations are qualitatively different from the HEOM exact
results. However, those comparisons fail in providing an overall intuitive picture about which model performs better as they are basis-dependent.
In other words, it is possible that one model may perform better or worse than another depending on the bases used. In this section, we
apply a measure of the non-Markovianity to develop a bases-free benchmark which can quantitatively describe the performance of the approximate methods.

\begin{figure}[h]
\includegraphics[width=\columnwidth]{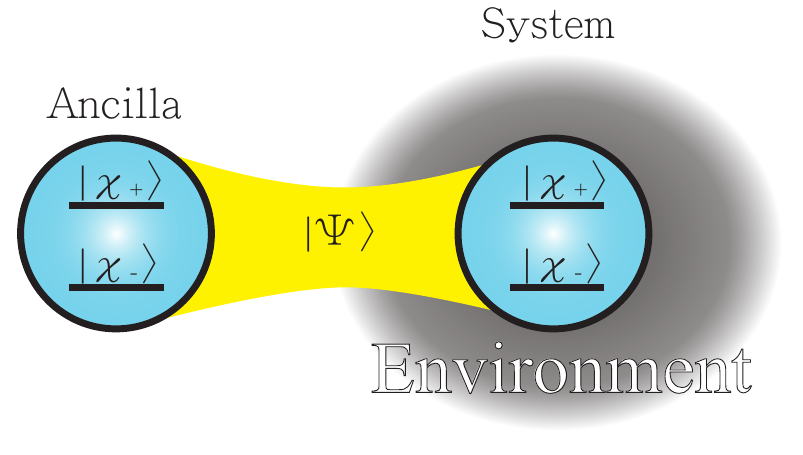}
\caption{(Color online) Schematic illustration of the entanglement measure. We consider a system and a copy of it acting as a well-isolated
ancilla possessing the same degrees of freedom of the system. Initially, they form a maximally-entangled state $\rho_{\textrm{sys},\textrm{anc}}(0)
=|\Psi\rangle\langle\Psi|$. Then the system starts to feel contact with its environment (denoted by the gray shadow) and evolves according
to $\mathcal{E}_{t,0}$, whereas the ancilla is kept isolated.}
\label{illustration}
\end{figure}

Let us consider an isolated ancilla possessing the same degrees of freedom of the system and with which the system forms a
maximally entangled initial state $|\Psi\rangle=\sum_{j=\chi_{+},\chi_{-}}\frac{1}{\sqrt{2}}|j\rangle\otimes|j\rangle$
(see Fig.~\ref{illustration}). If the system evolves according to a process $\mathcal{E}_{t,0}$, then the Choi-Jamio{\l}kowski isomorphism
\cite{jamiolkowski,choi} guarantees that the extended density matrix

\begin{eqnarray}
\rho_{\textrm{sys},\textrm{anc}}(t)&=&\left(\mathcal{E}_{t,0}\otimes\mathcal{I}_{\textrm{anc}}\right)\left(|\Psi\rangle\langle\Psi|\right)  \nonumber \\
&=&\sum_{j,k}\frac{1}{2} \mathcal{E}_{t,0}\left(|j\rangle \langle k|\right) \otimes |j\rangle \langle k| .
\end{eqnarray}
contains all the necessary information on the dynamics of the system, where $\mathcal{I}_{\textrm{anc}}$ is the identity process acting
on the ancilla. The entanglement, $\mathbf{E}(\rho_{\textrm{sys},\textrm{anc}})$, between the system and the ancilla is a physical quantity
which is typically very sensitive to environmental effects.

Another related quantity is the degree of non-Markovianity, $\mathcal{NM}$. Recently, many efforts have been devoted to construct a
proper measure of the non-Markovianity \cite{BLP_measure,RHB_measure}. Rivas \textit{et al.} \cite{RHB_measure} combine the concept of
the divisibility of a quantum process \cite{mmwolf1,mmwolf2} and the fact that no local completely positive (CP) operation \cite{choi}
can increase the entanglement $\mathbf{E}$ between a system and its corresponding ancilla
\begin{equation}
\mathbf{E}[\rho_{\textrm{sys},\textrm{anc}}]\geq\mathbf{E}\left[(\mathcal{E}_{\textrm{sys}}\otimes\mathcal{I}_{\textrm{anc}})(\rho_{\textrm{sys},\textrm{anc}})\right] .
\end{equation}
Consequently, Rivas \textit{et al.} \cite{RHB_measure} proposed that the degree of non-Markovianity within a given time interval
$[0,t]$ can be estimated by
\begin{equation}
\mathcal{NM}=\int_{0}^{t}\left|\frac{d}{d\tau}\mathbf{E}\left[\left(\mathcal{E}_{\tau,0}\otimes\mathcal{I}_{\textrm{anc}}\right)
\left(|\Psi\rangle\langle\Psi|\right)\right]\right|d\tau-\Delta\mathbf{E}_{t},
\label{RHP ent-measure}
\end{equation}
where
\begin{equation}
\Delta\mathbf{E}_{t}=\mathbf{E}\left[|\Psi\rangle\langle\Psi|\right]-
\mathbf{E}\left[\left(\mathcal{E}_{t,0}\otimes\mathcal{I}_{\textrm{anc}}\right)\left(|\Psi\rangle\langle\Psi|\right)\right].
\end{equation}
The non-Markovianity of open-system quantum dynamics can be evaluated at many different theoretical levels
\cite{BLP_measure,RHB_measure,NM_measure_exp,CM_nm_deg_2014,BCM_measure}, and the quantity $\mathcal{NM}$ is an extremely strict indicator
of non-Markovianity that measures the information exchange in time between the system and its environment. For $\mathcal{NM}$ to have a
non-zero value, explicit environmental memory effects must be present.

Here we compare the time evolution of the entanglement, $\mathbf{E}_{t}$, and the corresponding degree of non-Markovianity, $\mathcal{NM}$,
for the two approximate system-bath models and show how they can provide an integrated picture as to what extent their dynamics deviate
from the exact results.

\begin{figure*}[th]
\includegraphics[width=\textwidth]{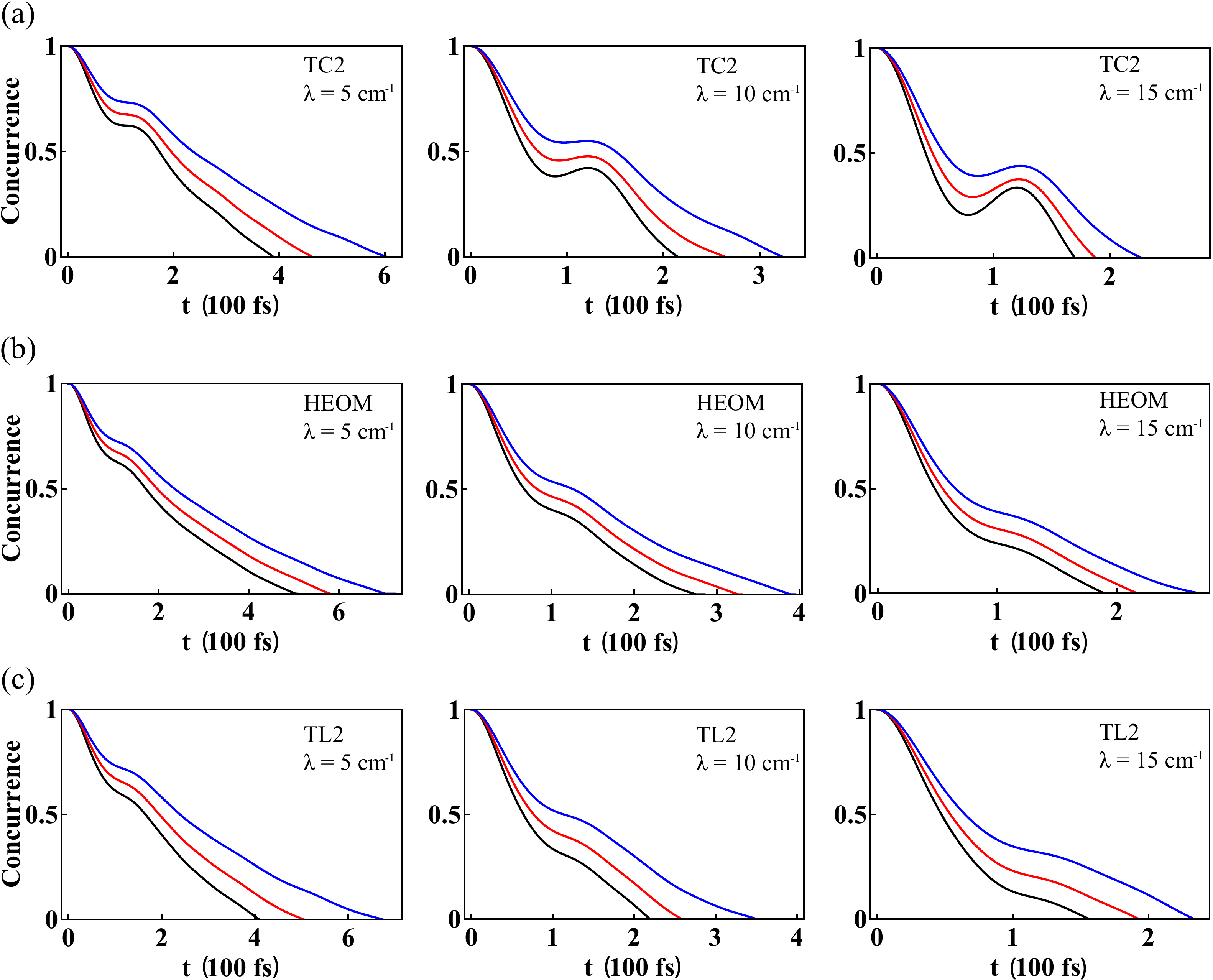}
\caption{(Color online) Time evolution of the concurrence calculated by (a) TC2, (b) HEOM, and (c) TL2 for different values of $\lambda$
at temperatures $T=300$ K (black), $250$ K (red), and $200$ K (blue). The other parameters are $\omega_{0}=70$ cm$^{-1}$, $J=100$ cm$^{-1}$,
and $\gamma=50$ cm$^{-1}$ ($\gamma^{-1}=106$ fs). In general, the concurrence will die out faster for larger $\lambda$ and higher temperatures.
The coherence over-estimation of the TC2 method is manifested by a concurrence revival around $100$ fs for larger values of $\lambda$,
whereas HEOM and TL2 produce a monotonically-decreasing concurrence.
}\label{Conc-all-50-70}
\end{figure*}

\subsection{Evaluating non-Markovianity}

\begin{figure}[th]
\includegraphics[width=\columnwidth]{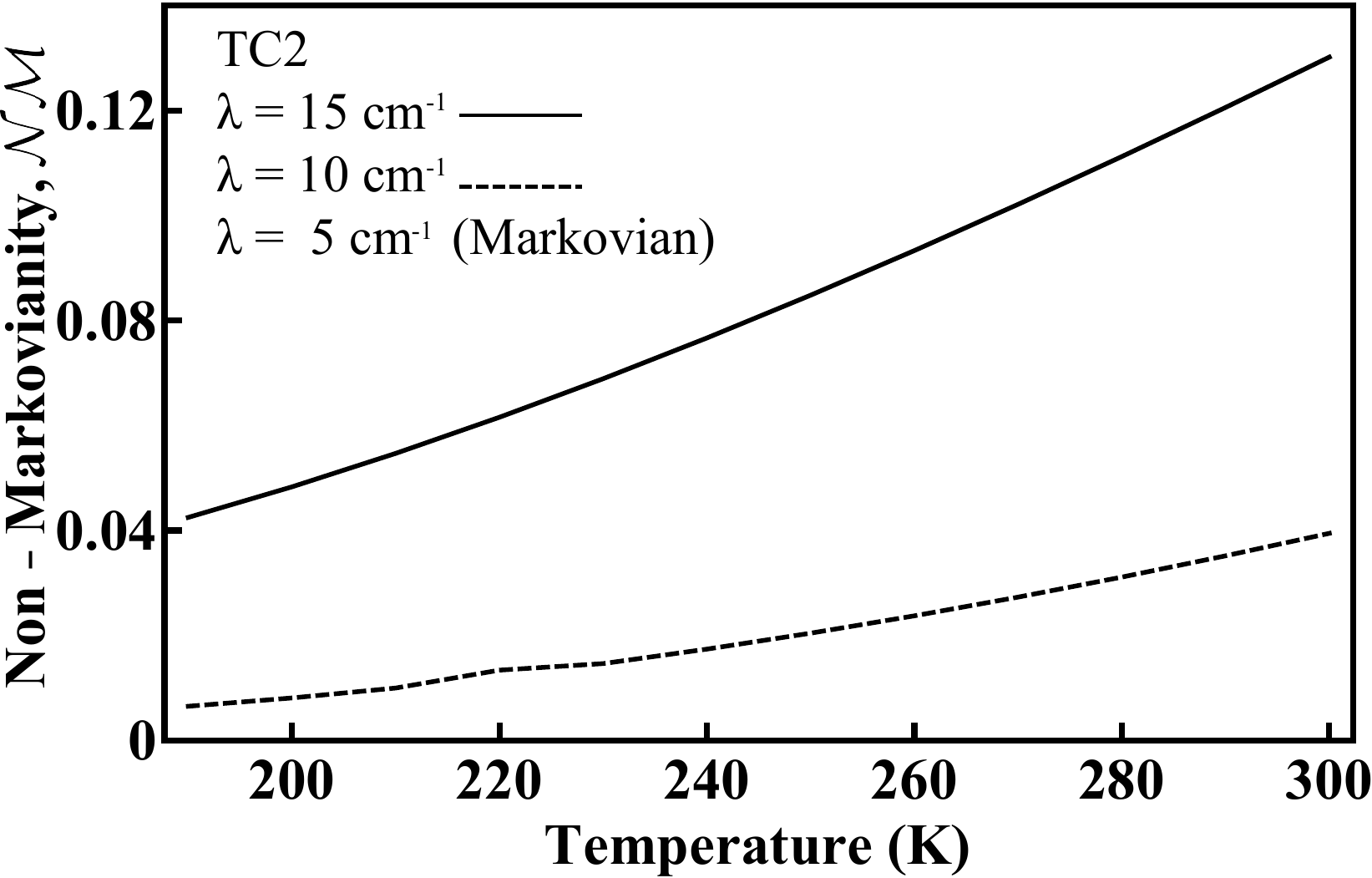}
\caption{Non-Markovianity, $\mathcal{NM}$, obtained by the TC2 method as a function of temperature. The parameters are the same as used in
Fig.~\ref{Conc-all-50-70}. Among the three methods investigated in this work, only TC2 at higher $\lambda$ generates non-zero non-Markovianity
for these parameters. At $\lambda=5$ cm$^{-1}$ TC2 correctly produces the expected Markovian dynamics for this regime.}
\label{Non-Marko-TC2-50-70}
\end{figure}

To analyse the behavior of the non-Markovianity in each method, in this section we will show how the concurrence,  a well-known measure
for bipartite entanglement \cite{wootters_concurrence}, between system and ancilla evolves in time and how the corresponding non-Markovianity
[Eq.~(\ref{RHP ent-measure})] depends on the physical parameters of the original spin-boson model.

As an explicit visualization of the integrand in Eq.~(\ref{RHP ent-measure}), in Fig.~\ref{Conc-all-50-70}, we apply the measure to
(a) TC2, (b) HEOM, and (c) TL2 and show the time evolution of the concurrence for different values of $\lambda$ at temperatures
$T=300$ K (black), $250$ K (red), and $200$ K (blue), respectively. The other parameters are $\omega_{0}=70$ cm$^{-1}$, $J=100$ cm$^{-1}$,
and $\gamma=50$ cm$^{-1}$ ($\gamma^{-1}=106$ fs). It can be seen that, when increasing the temperature and $\lambda $, the decoherence
becomes more pronounced. Hence, the concurrence will die out earlier for larger $\lambda$ and higher temperature. As shown in
Fig.~\ref{Conc-all-50-70}(a), except for $\lambda =5$ cm$^{-1}$, which produces monotonically-decreasing concurrence, the TC2 model produces
oscillatory curves, in which a concurrence revival is exhibited around $100$ fs and results in a finite degree of non-Markovianity (shown later).
While in Fig.~\ref{Conc-all-50-70}(b) and (c), HEOM and TL2 produce monotonically-decreasing concurrence and generate no visible non-Markovianity
with this measure.

In Fig.~\ref{Non-Marko-TC2-50-70}, we show the corresponding measure of the non-Markovianity, $\mathcal{NM}$, calculated using the time
evolution of the concurrence shown in Fig.~\ref{Conc-all-50-70}(a). Only TC2, for larger $\lambda $ values, leads to non-zero non-Markovianity,
while TC2 at $\lambda=5$ cm$^{-1}$,
HEOM, and TL2 generate null results due to the monotonically-decreasing concurrence. This comparison not only shows that the TL2 yields a better
approximation to the HEOM dynamics, but also explicitly demonstrates the degree to which TC2 deviates from HEOM. We again attribute this deviation
to the over-estimation of coherence shown in Fig.~\ref{Dynamics}. In addition, it can be seen in Fig.~\ref{Non-Marko-TC2-50-70} that $\mathcal{NM}$
tends to increase with increasing $\lambda$ and temperature. We will investigate this below in a regime where the HEOM results exhibit similar behavior.

\subsection{Increase of non-Markovianity with $\lambda$ and temperature}

The other two important parameters in our spin-boson model are the level spacing $\omega _{0}$ and the bath relaxation time $\gamma $. The
former affects to what extent the state $|\chi_{+}\rangle$ is delocalized, while the latter is related to the correlation time of the
environment and is directly connected to the non-Markovianity of the system.

\begin{figure}[ht]
\includegraphics[width=\columnwidth]{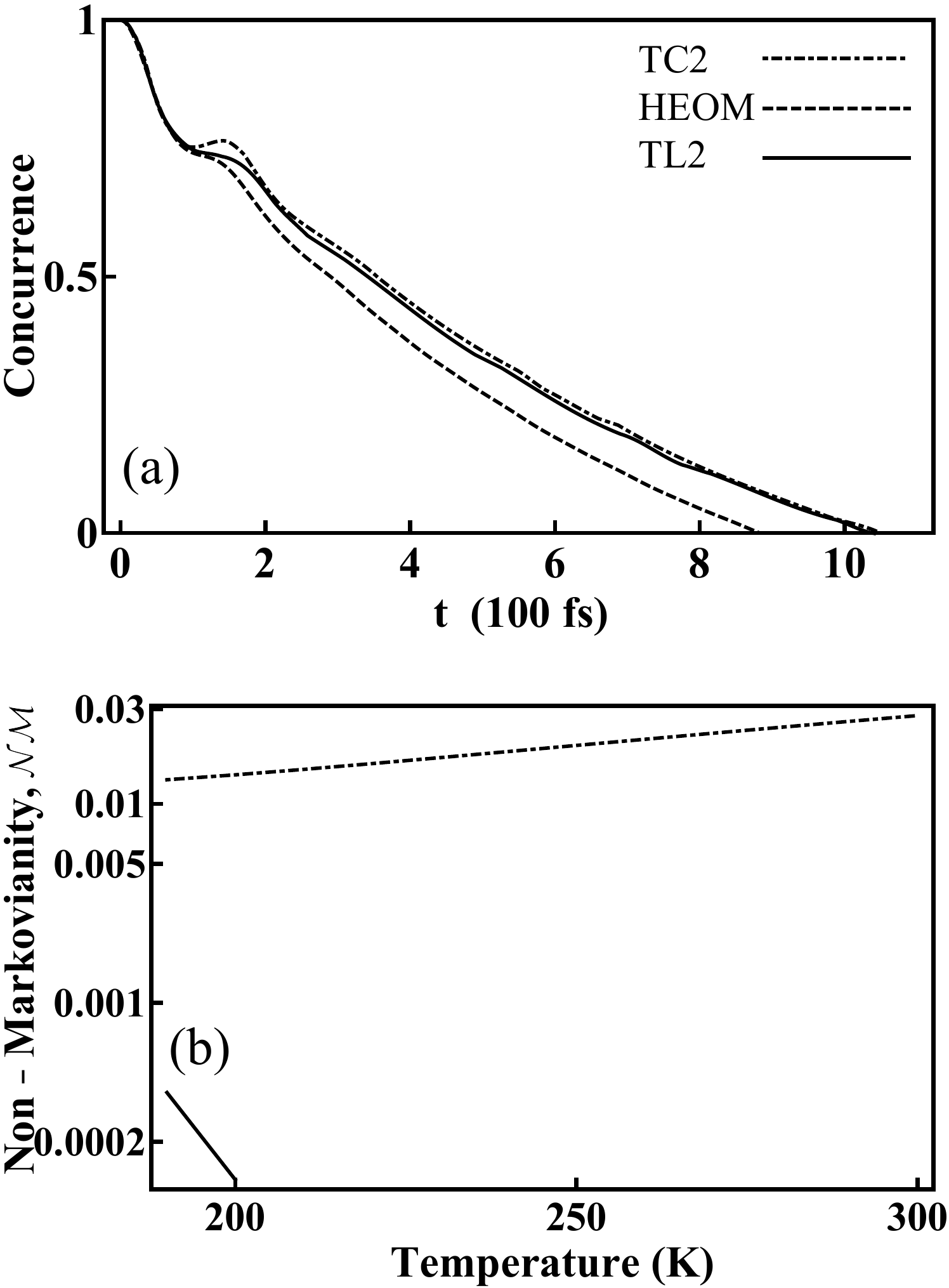}
\caption{(a) The concurrence obtained from the TC2 method (dot-dashed), HEOM (dashed), and TL2 (solid), for $\omega_{0}=40$ cm$^{-1}$. The
reduction of $\omega_{0}$ leads to a manifest concurrence revival around $100$ fs. The concurrence obtained from HEOM is still monotonically
decreasing. The other parameters are: $\lambda=5$ cm$^{-1}$, $\gamma=50$ cm$^{-1}$, and $T=200$ K. (b) The corresponding non-Markovianity
versus temperature. The result of TC2 shows finite non-Markovianity, while that from TL2 shows very small non-Markovianity and only at
low temperatures. The result from HEOM is Markovian due to its monotonically-decreasing concurrence.}
\label{Conc-all-50-05-40}
\end{figure}

In Fig.~\ref{Conc-all-50-05-40}(a), we reduce $\omega_{0}$ to $40$ cm$^{-1}$ and fix the other parameters at $\lambda=5$ cm$^{-1}$,
$\gamma=50$ cm$^{-1}$, and $T=200$ K. The reduction of $\omega _{0}$ leads to a manifest concurrence revival around $100$ fs in the TC2
concurrence dynamics,  a result of stronger delocalization and significant enhancement of the coherence effect. In the mean time,
the concurrence of the HEOM result is still monotonically decreasing. The TC2 model further over-estimates this enhancement and ends up
with finite non-Markovianity within all range of temperatures shown in Fig.~\ref{Conc-all-50-05-40}(b). The TL2 model predicts
almost-Markovian results, besides the very small non-Markovianity at low temperatures, again showing a better agreement with the HEOM exact
results.

\begin{figure}[]
\includegraphics[width=\columnwidth]{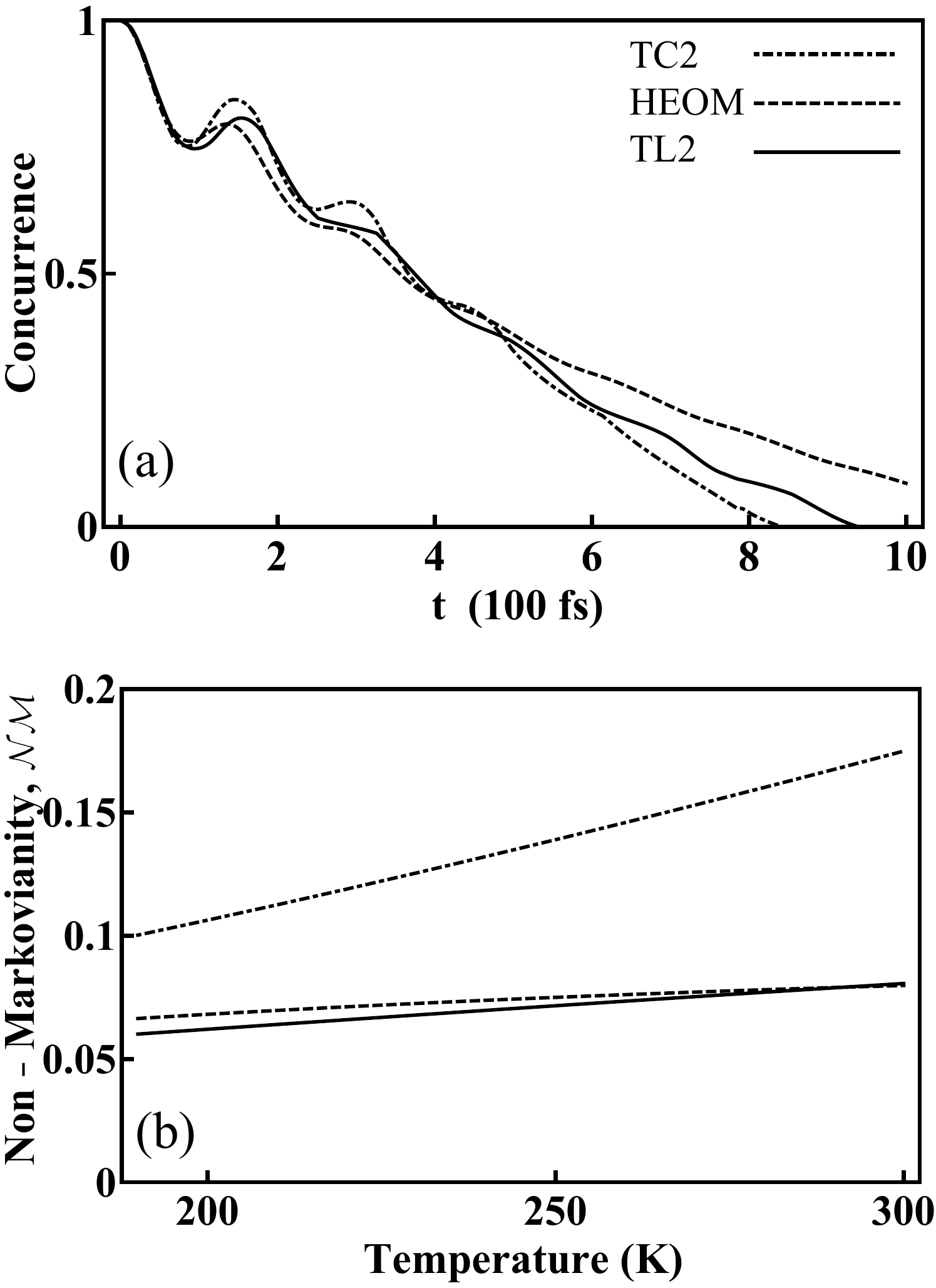}
\caption{(a) The concurrence versus time for TC2 (dot-dashed), HEOM (dashed), and TL2 (solid). The $\gamma$ value is further reduced to
$20$ cm$^{-1}$ ($\gamma^{-1}=265$ fs). The other parameters are the same as those
in Fig.~\ref{Conc-all-50-05-40}(a). The information on the system dynamics can possibly flow back from the environment into the system
and in turn leads to wavy concurrence curves. (b) The corresponding non-Markovianity values versus temperature. These non-Markovianity
values increase prominently as a result of the reduced $\gamma$ value. TC2 shows larger non-Markovianity values, while TL2 shows good
agreement with the HEOM.}
\label{Conc-all-20-05-40}
\end{figure}

In Fig.~\ref{Conc-all-20-05-40}(a), $\gamma$ is further reduced to $20$ cm$^{-1}$ ($\gamma^{-1}=265$ fs) to investigate the effect of
slow environments. As the spectral density function is narrower, the correlation time of the environment becomes long compared with the
characteristic time of the system dynamics. Hence the information on the system dynamics is more likely to be retained in the environment
and flow back into the system. This back-flow of information in turn affects the behavior of the system and results in beating in the
concurrence curves for all methods. As shown in Fig.~\ref{Conc-all-20-05-40}(b), the TC2 model predicts a non-Markovianity much larger
than the exact results. On the other hand, the TL2 model predicts a non-Markovianity in excellent agreement with the HEOM results, with
only a small under-estimation of the non-Markovianity in this set of parameters.

The above comparisons exhibit an interesting tendency for $\mathcal{NM}$ to increase with $\lambda$ and temperature. Several relevant
theoretical and experimental works have reported \cite{nazir_prl_2009,fleming_science_2007,scholes_nature_2010} that strong
system-environment correlations are helpful for maintaining quantum coherence even at high temperatures. As a result, higher
temperature may in turn activate more phonon modes in the environment without destroying the quantum coherence significantly. This provides
more channels via which the system can interact with the environment. In the language of quantum information science, smaller $\gamma$ and
strong system-environment correlation may help to preserve the dynamical information; while larger $\lambda$ and higher temperature may increase
the possibility that this information can flow back into the system from the environment. Consequently, this increase of $\mathcal{NM}$
with larger temperature and $\lambda$ is a result of the competition between the back-flow of information and thermal fluctuations.
Meanwhile, the magnitude of the concurrence is reduced by the stronger random fluctuations in the environment.

\section{CONCLUSIONS}

In summary, we first investigate the dynamics of two perturbative second-order QME methods, TC2 and TL2, and compare their results with
the numerically-exact results calculated by HEOM. We find that TC2 can approximate the HEOM population better than TL2. However, a drawback
of the TC2 model is its over-estimation of the coherence. This drawback results in the TC2 model predicting too much beating behavior in
the population dynamics and limits the accuracy of TC2. In constrast, the TL2 model predicts  sluggish dynamics and loss of coherence
faster than that of the exact HEOM. As a result, the population tends to reach thermal equilibrium too rapidly.

To further investigate the dynamics and establish a benchmark for the performance of perturbative QMEs, we combine the concept of
Choi-Jamio{\l}kowski isomorphism \cite{jamiolkowski,choi}, entanglement with an ancilla \cite{wootters_concurrence}, and a measure of
non-Markovianity \cite{RHB_measure} to provide a quantitative way to determine how much the coherence dynamics and memory effects are
deviating from the exact result. This provides a deep physical insight on the effects of each parameter and a single quantity to determine
how much the QME dynamics deviates from the exact results. Here we find that the non-Markovian measure indicates that the TL2 approximates
HEOM better than TC2 in terms of the coherence dynamics and memory effects for the dimer system studied here. In addition, while it is well
understood that the reorganization energy $\lambda$ and temperature enhance the effect of thermal fluctuations in the environment on the
system, increasing these parameters can have surprising results. In particular, our results show that higher temperature increases information
back-flow from the environment, thus increasing the non-Markovianity of the system dynamics, even though the concurrence itself undergoes faster
decay. These results could have important implications in the theoretical modeling of electronic coherence in photosynthetic
systems \cite{yuanchun_review_article,yuehnan_nat_phy_2013,pachon_pccp_2012}.

\section*{}

\section*{ACKNOWLEDGMENTS}

This work is supported partially by the National Center for Theoretical Sciences and Minister of Science and Technology, Taiwan, grant numbers
NSC 101-2628-M-006-003-MY3, and MOST 103-2112-M-006 -017 -MY4. FN is partially supported by the RIKEN iTHES Project, MURI Center for Dynamic
Magneto-Optics, and a Grant-in-Aid for Scientific Research (S). YCC thanks the Ministry of Science and Technology, Taiwan (Grant No.
NSC 100-2113-M-002-008-MY3).

\begin{widetext}

\section*{APPENDIX: FULL EXPRESSIONS FOR THE TC2 AND TL2 QUANTUM MASTER EQUATIONS}

The detailed expression of the TC2 integrodifferential QME Eq.~(\ref{int_diff_TC2}) is given by
\begin{eqnarray}
\frac{\partial}{\partial t}\left[e^{-i\omega_{\alpha,\beta}t}\rho_{\alpha,\beta}(t)\right]&=&
-i\omega_{\alpha,\beta}\exp\left[-i\omega_{\alpha,\beta}t\right]\rho_{\alpha,\beta}(t)   \nonumber\\
&&+\sum_{\mu,\nu}A_{\alpha,\nu}A_{\mu,\beta}\int_{0}^{t}G(t-\tau)e^{i\omega_{\mu,\alpha}(t-\tau)}e^{-i\omega_{\nu,\mu}\tau}\rho_{\nu,\mu}(\tau)d\tau\nonumber\\
&&-\sum_{\mu,\nu}A_{\alpha,\nu}A_{\nu,\mu}\int_{0}^{t}G(t-\tau)e^{i\omega_{\beta,\nu}(t-\tau)}e^{-i\omega_{\mu,\beta}\tau}\rho_{\mu,\beta}(\tau)d\tau\nonumber\\
&&+\sum_{\mu,\nu}A_{\alpha,\nu}A_{\mu,\beta}\int_{0}^{t}G^{*}(t-\tau)e^{i\omega_{\beta,\nu}(t-\tau)}e^{-i\omega_{\nu,\mu}\tau}\rho_{\nu,\mu}(\tau)d\tau\nonumber\\
&&-\sum_{\mu,\nu}A_{\nu,\mu}A_{\mu,\beta}\int_{0}^{t}G^{*}(t-\tau)e^{i\omega_{\mu,\alpha}(t-\tau)}e^{-i\omega_{\alpha,\nu}\tau}\rho_{\alpha,\nu}(\tau)d\tau,\nonumber\\
\end{eqnarray}
where $G(t)$ is the correlation function defined by Eq.~(\ref{correlation_function}). Whereas the detailed expression of TL2
QME in Eq.~(\ref{int_diff_TL2}) is given by
\begin{eqnarray}
\frac{\partial}{\partial t}\left[e^{-i\omega_{\alpha,\beta}t}\rho_{\alpha,\beta}(t)\right]&=&
-i\omega_{\alpha,\beta}\exp\left[-i\omega_{\alpha,\beta}t\right]\rho_{\alpha,\beta}(t)   \nonumber\\
&&+\sum_{\mu,\nu}A_{\alpha,\nu}A_{\mu,\beta}\int_{0}^{t}G(t-\tau)e^{i\omega_{\nu,\alpha}(t-\tau)}e^{-i\omega_{\nu,\mu}t}\rho_{\nu,\mu}(t)d\tau\nonumber\\
&&-\sum_{\mu,\nu}A_{\alpha,\nu}A_{\nu,\mu}\int_{0}^{t}G(t-\tau)e^{i\omega_{\mu,\nu}(t-\tau)}e^{-i\omega_{\mu,\beta}t}\rho_{\mu,\beta}(t)d\tau\nonumber\\
&&+\sum_{\mu,\nu}A_{\alpha,\nu}A_{\mu,\beta}\int_{0}^{t}G^{*}(t-\tau)e^{i\omega_{\beta,\mu}(t-\tau)}e^{-i\omega_{\nu,\mu}t}\rho_{\nu,\mu}(t)d\tau\nonumber\\
&&-\sum_{\mu,\nu}A_{\nu,\mu}A_{\mu,\beta}\int_{0}^{t}G^{*}(t-\tau)e^{i\omega_{\mu,\nu}(t-\tau)}e^{-i\omega_{\alpha,\nu}t}\rho_{\alpha,\nu}(t)d\tau.\nonumber\\
\end{eqnarray}
\end{widetext}

\bibliographystyle{apsrev4-1}
\bibliography{non_Marko_bench}

\end{document}